\documentclass{PoS}

\title{Energetic axion-like particle production in galaxies}

\ShortTitle{Energetic axion-like particle production in galaxies}

\author{\speaker{Athanasios Manousos}\\
        School of Physics\\
				Nuclear and Elementary Particle Physics Department\\
				Aristotle University of Thessaloniki\\
				GR-54124, Thessaloniki, Greece\\
        E-mail: \email{athmanou@auth.gr}}

\author{Anastasios Liolios\\
        School of Physics\\
				Nuclear and Elementary Particle Physics Department\\
				Aristotle University of Thessaloniki\\
				GR-54124, Thessaloniki, Greece\\
        E-mail: \email{lioliosa@auth.gr}}
        
\author{Christos Eleftheriadis\\
        School of Physics\\
				Nuclear and Elementary Particle Physics Department\\
				Aristotle University of Thessaloniki\\
				GR-54124, Thessaloniki, Greece\\
        E-mail: \email{christos.eleftheriadis@cern.ch}}

\abstract{Relativistic axion-like particles (ALPs) originating from the stellar interiors, along with the ones coming from photon-ALP mixing in the galactic magnetic fields, contribute together to make an energetic component of the ALP content of the universe. Considering an isotropic distribution of cosmic gamma rays, it is examined if these high energy axion-like particles could constitute a measurable isotropic "background" for helioscope-type and existing large general-purpose particle experiments, taking into account their sensitivity.}

\FullConference{Identification of Dark Matter 2010-IDM2010\\
		July 26-30, 2010\\
		Montpellier France}

\begin{document}

\section{Introduction}
Axions are hypothetical pseudo-scalar particles whose existence has been proposed as a solution to the so-called "strong CP problem" \cite{KC}. Today they constitute one of the most promising candidates for the dark matter of the universe \cite{Duffy}. Recently their existence has been related with a more general class of particles, the so-called "axion-like particles" (ALPs) \cite{AXIONS},\cite{Z-NJP}. ALPs are scalar or pseudo-scalar particles of very low mass like the axions. Their coupling to photons is not connected to their mass, as is the case for QCD-axions. Their origin may be either cosmological or astrophysical, mainly through photon-ALP conversion under the influence of strong electric and magnetic fields. Cosmological ALPs are of very low mass and energy and they are considered to be part of the cold dark matter in the cosmos \cite{Asztalos},\cite{Duffy},\cite{Serpico}. The higher their mass the lower their contribution in the dark matter content of the universe.

ALPs may be produced from photon conversion through the Primakoff effect \cite{Primakoff}. This may happen, when a photon flux crosses strong electric and magnetic fields. Such a case may exist in stellar interiors, where thermal photons are scattered under the influence of strong plasma fields. The ALP flux expected from sun-like stars has been calculated and ALP spectra are given as a function of energy \cite{AXIONS}. Photon-ALP conversion in large-scale magnetic fields and the existence of ALPs which have a strong coupling to two photons have led to a series of publications in the last decade investigating a number of interesting astrophysical consequences \cite{DARMA-S},\cite{DARMA-ALP},\cite{Paneque-ALP},\cite{Hooper},\cite{DARMA-AS}. For example, photon to ALP conversion might explain the dimming of distant very high energy gamma-ray sources \cite{Grossman-Ia},\cite{Csaki},\cite{AXIONS} since it is expected that in the presence of magnetic fields at sufficiently long distances, the conversion effect saturates and, in average, one third of high energy photons are converted to ALPs \cite{Burrage-2009}. The dimming of the light from distant supernovae has also been discussed in the context of photon-ALP conversion, as an alternative to the accelerated expansion of the universe \cite{Csaki},\cite{Mirizzi-PA}.

In the present work it is examined if these highly relativistic ALPs could constitute a measurable isotropic "background" for helioscope-type experiments, taking into account their sensitivity. We also examine if some of these energetic ALPs could be detected using particle-physics detectors like the ones in the ATLAS and CMS experiments at LHC.

\section{ALP-production in galaxies}
It is interesting to consider the production of stellar ALPs in the Galaxy, as well as the ALP production in extragalactic powerful objects like AGNs and GRBs. In the next section we shall examine the expected ALP fluxes on earth from these objects due to photon-ALP conversion in cosmic long-scale magnetic domains. 

Thermal photons produced in the star interiors can be transformed into ALPs with the same energy in the fluctuating electromagnetic fields of the stellar plasma. In the case of our sun, the solar axion luminosity has been calculated \cite{AXIONS}. The spectrum is continuous with the maximum of the distribution at $3.0\:keV$ and average energy of $4.2\:keV$. The integrated solar axion luminosity is:
\begin{equation}
	L_{\alpha}=\left(g_{\alpha\gamma\gamma}/10^{-10}GeV^{-1}\right)^2\times1.85\times10^{-3}L_{SUN}
\label{eq:ISL}
\end{equation}
Taking a solar X-ray luminosity of $10^{21} J s^{-1}$ and $g_{\alpha\gamma\gamma}\approx10^{-10}GeV^{-1}$ as representative values for $1-10\:keV$ photons, we have a solar axion luminosity of the order $10^{18} J s^{-1}$. Since in our Galaxy $10^{11}$ to $10^{12}$ stars exist, we can make the rough estimation that the ALP galactic stellar production in the $1-10\:keV$ region might be of the order $10^{30} J s^{-1}$. 

ALP luminosity of Active Galactic Nuclei can be estimated within the scenario that ALPs can be produced in the accretion disk by the Compton, bremsstrahlung and Primakoff processes. For bremsstrahlung which is the most efficient mechanism, this luminosity has been calculated and found to be of the order $10^{29} J s^{-1}$ \cite{Jain-AL}. Fairbairn et al. \cite{Fairbairn-LAB} have suggested that a flux of ALPs on earth is expected coming from conversion of energetic gamma-rays emitted by astrophysical sources to ALPs in the magnetic fields of the sources themselves. In their work they estimate the contribution of various astrophysical sources to the ALP flux from photons with energies $E>10\:keV$ (mainly pulsars and gamma-ray bursts). For other sources, like AGNs, there is a contribution only at high energies ($E>10\:MeV$) where the fluxes are considerably lower. The expected flux of astrophysical ALPs is estimated to be of the order $~10^{-3} cm^{-2} s^{-1} MeV^{-1}$ at about $1\:MeV$, and $~10^{-5} cm^{-2} s^{-1} MeV^{-1}$ at about $10\:MeV$ \cite{Fairbairn-LAB}. The expected ALP flux is decreasing for higher energies with a spectral index of about 2 - 2.3. Extrapolating, we can estimate a flux of about $~10^{-9} cm^{-2} s^{-1} MeV^{-1}$ at about $1\:GeV$.

\section{Photon to ALP conversion in cosmic magnetic domains}
In the presence of large-scale magnetic fields, photons could convert to ALPs and vice versa. This ALP-photon oscillation could give rise to a flux of energetic ALPs. The mixing depends on the photon energy E and a number of other independent parameters like the plasma frequency of the propagating medium, the ALP rest mass and the characteristics of the intergalactic magnetic domains. The most interesting case is the possibility of strong mixing between photons and ALPs in which the effect is enhanced and saturates after traversing large intergalactic distances.

The dependence of the probability of the ALP-photon mixing on the energy, $P=P\left(E\right)$ \cite{Burrage-2009},\cite{Pettinari-2010}, can be written as:
\begin{equation}
 P\left(E\right)=A\left(E\right)\sin^2(\frac{gBL}{2\sqrt{A\left(E\right)}})
\label{eq:Probability}
\end{equation}
where B and L are the mean magnetic field and the characteristic length, respectively, in the cosmic magnetic regions, g is the photon-ALP coupling constant, and $A\left(E\right)=\frac{E^{2}}{E^{2}+E^{2}_{c}}\in\left(0,1\right)$, is a factor depending on the energy E of the particle and a characteristic energy scale $E_{c}$ determined by parameters like the ALP effective mass, the photon-ALP coupling strength, etc.
\begin{figure}[htbp]
	\centering
		\includegraphics[width=.6\textwidth]{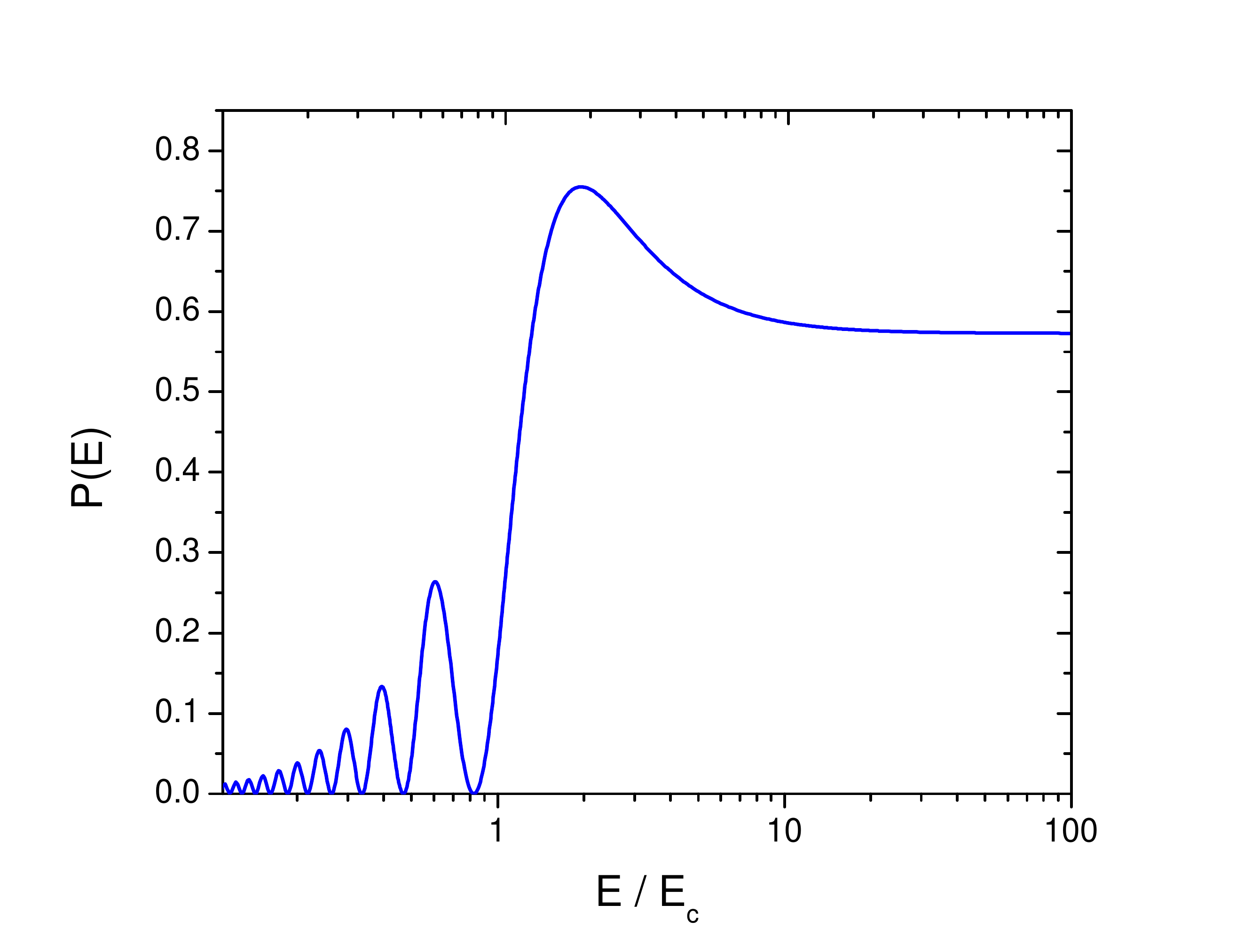}
	\caption{Dependence of the ALP-photon mixing probability on the energy $\left(E/E_{c}\right)$}
	\label{fig:Fig1}
\end{figure}
For the behaviour of the above probability (3.1) an example is given in Figure 1. For energies $E/E_{c}>10$ the mixing is relatively strong and independent on the photon energy. Burrage, Davis and Shaw \cite{Burrage-2009} estimate that mixing is most significant for X and $\gamma$-rays since the strong mixing limit is reached when $E\geq0.3-3\:keV$. Also, they expect that one third of the gamma-ray photons on average are converted to ALPs when a variety of photons paths cross a large number of randomly oriented magnetic regions.

Here, we shall focus on the so-called "extragalactic diffuse $\gamma$-ray background" (EGB) \cite{FERMI},\cite{Chen} and the extraction of an associated ALP flux. The Fermi LAT Collaboration found that the spectrum of the EGB is consistent with a power law with a differential spectral index of $~2.4$ and intensity (for $E>100\:MeV$) $~10^{-5} cm^{-2} s^{-1} sr^{-1}$. Assuming that the observed EGB-spectrum is what remains from an initial photon flux after its transformation by one third into an ALP-flux, we estimate that an "extragalactic diffuse ALP emission" exists, with intensity $~5\times10^{-6} cm^{-2} s^{-1} sr^{-1}$ for energies above $100\:MeV$. The ALP spectrum is expected to be similar to the gamma ray spectrum, since the ALP-photon mixing probability is independent on the energy above the strong mixing limit $(E>10E_{c})$. The characteristic energy $E_{c}$ can take values in a very broad range of energies depending on the assumptions \cite{Serpico},\cite{Paneque-ALP},\cite{Hooper},\cite{DARMA-AS},\cite{Burrage-2009}. For ultralight ALPs $(m_{a}\approx10^{-11} eV)$, it could be in the $MeV$ region for typical astrophysical parameter values. This assumption is optimistic of course, but it introduces a bound in the flux that one would expect as an "ALP-background" in axion searches (Figure 2).
\begin{figure}[htbp]
	\centering
		\includegraphics[width=.6\textwidth]{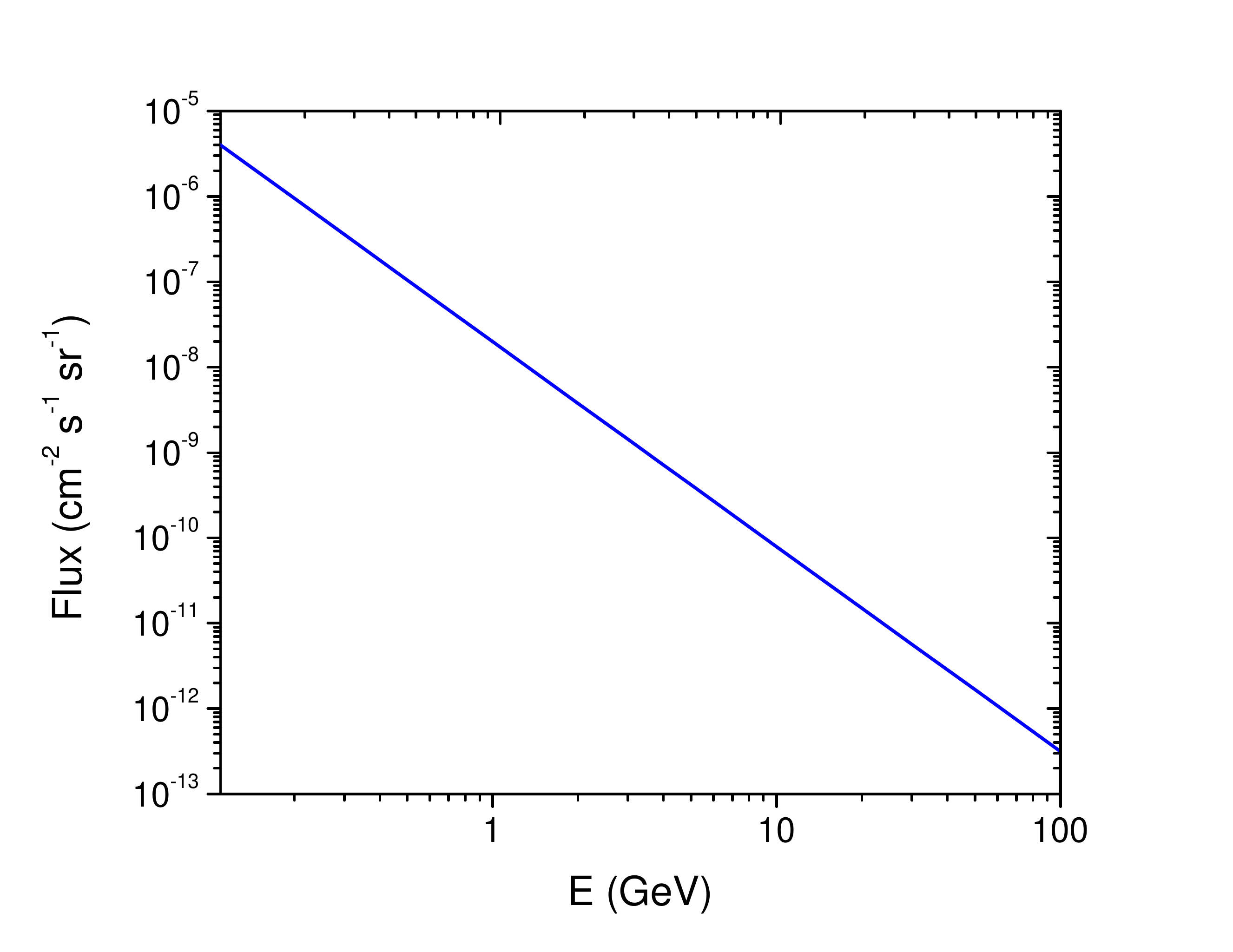}
	\caption{Estimated flux of the extragalactic diffuse ALP background. The spectrum follows a power law with a spectral index of $-~2.4$}
	\label{fig:Fig2}
\end{figure}

\section{The energetic "ALP-background" in axionscopes and in big particle detectors}
As mentioned in the previous two sections, the energetic $(E>1\:MeV$) ALP production in the Galaxy is expected to produce a flux lower or, in the best case, of the same order as the flux from photon - ALP conversion in the intergalactic magnetic fields. The point now is to find a way to check the hypothesis for diffuse energetic ALP flux, by exploiting existing detector equipment. Helioscopes like CAST \cite{Z-CAST},\cite{celefthe},\cite{Ruz-CAST} and the Tokyo Axion Helioscope \cite{Moriyama},\cite{Inoue-PL},\cite{Inoue-POS} are designed for detecting solar axions mainly in the energy range up to $100\:keV$. CAST experiment has also performed measurements in the energy range up to $70\:MeV$, searching for solar axion emission lines from nuclear M1 transitions \cite{CAST-2010}. The background in CAST's low-background gamma-ray calorimeter was of the order $10^{-3}$ to $10^{-4} cm^{-2} s^{-1}$ in the energy range $0.3$ to $70\:MeV$, while no excess signal found above background at $0.478\:MeV$ and $5.5\:MeV$ axion lines. On the other hand, there is the possibility for terrestrial ALP production ("geo-axions") due to nuclear disintegrations of natural radionuclides in Earth's interior \cite{Liolios}. An optimistic estimate of their flux on Earth's surface give about $10^{-1} cm^{-2} s^{-1}$ which is not detectable by axionscopes. Since the expected solar ALP signal is of many orders of magnitude stronger than that from diffuse energetic ALP-background, there is no chance for the later to be detected in such experiments.

It is interesting to examine also the possibility for detecting a potential signal from diffuse energetic ALP flux, by exploiting existing detector equipment of the big detector systems of nowadays, such as ATLAS \cite{ATLAS} and CMS \cite{CMS}. The equipment which is most promising for detection of single photon events at energies above $~10\:GeV$ is the superconducting solenoids of these multi-purpose detectors with their meter-scale size and Tesla-scale magnetic fields. A high energy cosmic ALP crossing the ATLAS solenoid encounters a magnetic field of 2 Tesla. The ALP direction has in general a component vertical to the direction of the magnetic field. This condition applies also for some areas of the toroidal magnet system. On the other hand, in CMS the solenoid is bigger, 14 meters in diameter, and the magnetic field is stronger, i.e. 4 Tesla. 

With an ALP flux of $10^{-10} cm^{-2} s^{-1}$, a number of $10^{-4} s^{-1}$ is reasonable to cross the solenoid (or 10 ALPs per day) at energies above $~10\:GeV$. The probability of producing a photon from an energetic ALP \cite{Z-OLD} entering a magnetic field with the above characteristics can be in the best case of the order $10^{-18}$. Thus, the expected count-rate due to the above ALP flux is estimated to be less than $10^{-14}$ counts per year. Obviously, it is not possible to detect an ALP signal also in LHC large detector systems.

\end{document}